\documentstyle[prl,aps,multicol,epsf]{revtex}
\renewcommand{\narrowtext}{\begin{multicols}{2} \global\columnwidth20.5pc}
\renewcommand{\widetext}{\end{multicols} \global\columnwidth42.5pc} 

\begin{document}

\newcommand{\be}{\begin{equation}}
\newcommand{\ee}{\end{equation}}
\newcommand{\bea}{\begin{eqnarray}}
\newcommand{\eea}{\end{eqnarray}}
\newcommand{\nt}{\narrowtext}
\newcommand{\wt}{\widetext}

\title{Magnetic field-induced insulating behavior in highly oriented pyrolitic
graphite}

\author{D. V. Khveshchenko}

\address{Department of Physics and Astronomy, University of North
Carolina, Chapel Hill, NC 27599}
\maketitle

\begin{abstract}
We propose an explanation for the apparent semimetal-insulator transition observed 
in highly oriented pyrolitic graphite in the presence of magnetic field perpendicular to the layers.
We show that the magnetic field opens an excitonic gap in the linear spectrum
of the Coulomb interacting quasiparticles, in a close analogy with the phenomenon of  
dynamical chiral symmetry breaking in the relativistic theories of the $2+1$-dimensional 
Dirac fermions. Our strong-coupling appoach allows for a non-perturbative 
description of the corresponding critical behavior. 
\end{abstract}
\nt

The recently discovered carbon-based materials provide a new playground for  
applications of the advanced methods of quantum field theory. 
Although the best known example is that of the one-dimensional 
carbon nanotubes described as the Luttinger liquid, 
some of the available non-perturbative techniques can also
be applied to higher dimensional
systems, such as layered highly oriented pyrolitic graphite (HOPG). 

In a single sheet of graphite, the low-energy spectrum of the  
quasiparticle excitations becomes linear in the vicinity of the two 
conical points located at the opposite corners of the two-dimensional  
Brillouin zone where the conduction and valence bands touch upon each other \cite{Semenoff}.
These low-energy excitations can be conveniently  
described in terms of a four-component Dirac spinor 
$\Psi_\sigma=(\psi_{1A\sigma}, \psi_{1B\sigma}, \psi_{2A\sigma}, \psi_{2B\sigma})$,
combining the Bloch states $\psi_{i\sigma}({\bf r})$ with spin $\sigma$ 
which are composed of the momenta near one of the 
conical points $(i=1,2)$ and propagate independently on the two different  
sublattices $({\bf r}=A,B)$ of the bi-partite hexagonal lattice of the graphite sheet.
In the following discussion, we will treat the number of the spin components $N$ 
as a tunable parameter, the physical case corresponding to $N=2$.

The use of the Dirac spinor representation
allows one to cast the quasiparticle kinetic energy in the relativistic-like form
\be
K=iv\sum_{\sigma =1}^N \int d^2{\bf r}{\overline \Psi}_\sigma
({\hat \gamma}_1\nabla_x+{\hat \gamma}_2\nabla_y)\Psi_\sigma
\ee
where ${\overline \Psi}_\sigma=\Psi^\dagger_\sigma{\gamma}_0$.
The reducible representation of the $4\times 4$ $\hat \gamma$-matrices 
${\hat \gamma}_{0,1,2}=(\tau_{3}, i\tau_{2}, -i\tau_{1})\otimes\tau_3$ 
satisfying the usual anticommutation relations  
$\{{\hat \gamma}_{\mu},{\hat \gamma}_{\nu}\}=2 {\rm diag}(1, -1, -1)$ is  
given in terms of the triplet of the Pauli matrices $\tau_{i}$,
and the velocity $v\sim 10^6m/s$ is proportional to the width of the 
electronic $\pi$-orbital band.

The Lorentz invariance of the non-interacting Hamiltonian is not, however,  
respected by the interaction term 
$$
U=
{g\over 4\pi}\sum^{N}_{\sigma,\sigma^\prime=1}\int {d^2{\bf r}d^2{\bf r}^\prime}
$$
\be
{\overline \Psi}_{\sigma}({\bf r}){\hat \gamma}_0\Psi_{\sigma}
({\bf r}){1\over {|{\bf r}-{\bf r}^\prime|}}
{\overline \Psi}_{\sigma^\prime}({\bf r}^\prime)
{\hat \gamma}_0\Psi_{\sigma^\prime}({\bf r}^\prime)
\ee 
which accounts for the long-ranged part of the Coulomb coupling
whose strength is characterized by the dimensionless parameter $g=2\pi e^2/\epsilon_0v$. 

The earlier perturbative studies of the effects of the Coulomb interaction                                                                              
resulted in the prediction that, upon renormalization, the strength of the 
effective coupling 
$g(\epsilon)\sim 1/|\ln\epsilon|$ monotonically decreases with the energy $\epsilon$, 
hence the paramagnetic semimetallic ground state remains stable \cite{Guinea}.
However, this conclusion appears to contradict the recent experimental observation
of a ferromagnetic bulk magnetization (inconsistent with the estimated number of 
magnetic impurities) in the HOPG samples showing insulating behavior of the resistivity 
\cite{Kopelevich}. 

In fact, the large value of the bare Coulomb coupling $g\gtrsim 10$ suggests that 
a more appropriate starting point might the strong coupling regime
where perturbation theory fails and a more capable approach is needed. 
In the present paper, we propose such an approach 
by focussing on a recent experimental observation of the apparent
magnetic field-driven semimetal-insulator transition in HOPG \cite{Kempa} 
and demonstrate that external magnetic field can trigger the instability 
towards excitonic insulator phase. 
Interestingly enough, the latter appears to have much in common with
the phenomenon of chiral symmetry breaking (CSB) which has been previously studied 
in the relativistic theories of the interacting Dirac fermions. 

In the case of graphite, the issue of CSB comes about due to 
the invariance of Eqs.(1,2) under arbitrary $U(2N)$ rotations
of the $2N$-component vector comprised of the chiral Dirac fermions  
$\Psi_{(L,R)\sigma}={1\over 2}({\bf 1}\pm{\hat \gamma}_5)\Psi_\sigma$, 
where the matrix ${\hat \gamma}_5=
{\bf 1}\otimes{\tau_2}$ anticommutes with any ${\hat \gamma}_\mu$. 

In a quantum system, strong interactions can give rise
to the appearance of a fermion mass and gapping of the fermion spectrum, 
thereby breaking the continuous chiral symmetry $U(2N)$ down to its subgroup
$U(N)\otimes U(N)$ which corresponds to the independent rotations of 
$\Psi_{L\sigma}$ and $\Psi_{R\sigma}$.

As one important $2+1$-dimensional example, CSB can be caused by the interaction
with a scalar Higgs-Yukawa (HY) bosonic mode coupled to the Dirac fermions
via the mass operator $\sum^N_{\sigma}{\overline\Psi}_\sigma\Psi_\sigma$.
In this case, CSB is known to 
occur for any number of fermion species $N$, provided that the strength of the  
(intrinsically attractive) HY coupling exceeds a certain critical value. 
Recently, this model has been applied to the analysis of the condensed matter
systems where the scalar bosonic field describes fluctuations of
a superconducting order parameter \cite{Jens} or piesoelectric phonons \cite{Antonio}.

In contrast, the repulsive Lorentz-invariant vector-like 
coupling via the current operator $\sum^N_\sigma{\overline\Psi}{\hat \gamma}_\mu\Psi$
drives the Dirac fermions towards the CSB
transition, regardless of the coupling strength,  
provided that the number of fermion species $N$ is sufficiently small ($N<N_c$).  
This behavior is believed to occur in the strong coupling infrared fixed point in the
$2+1$-dimensional Quantum Electrodynamics ($QED_3$) where the zero temperature value
of $N_c$ was found to be smaller or equal to $3/2$ \cite{Appelquist}.

However, this situation
changes drastically in the presence of magnetic field
which suppresses the orbital motion of the 
Dirac fermions and collapses their spectrum into a discrete set of the (relativistic) 
Landau levels. This effectively reduces the dimensionality of the problem, thus
enabling CSB to occur, regardless of the coupling strength and/or the 
number $N$ of fermion species.

In the Dirac picture of the quasiparticle excitations in layered graphite,  
the onset of CSB would be manifested by a non-zero value of the  
order parameter $\sum_\sigma{\overline \Psi}_\sigma \Psi_\sigma
=\sum_{i\sigma}(\psi^\dagger_{iA\sigma}\psi_{iA\sigma}-
\psi^\dagger_{iB\sigma}\psi_{iB\sigma})$. It
determines the magnitude of
the fermion gap proportional to the 
electron density imbalance between the $A$ and $B$ sublattices and corresponds to the formation of a  
site-centered charge density wave (CDW) in the excitonic insulating
ground state. 

While for a sufficiently small $N$ 
the excitonic instability could develop even in a single layer of graphite \cite{Leal},  
in the physical case $N=2$ it is unlikely to occur 
in the absence of the inter-layer Coulomb repulsion.
Indeed, in a realistic HOPG system consisting of many layers stacked in a staggered configuration,
the latter favors spontaneous depletion of the electron density on one of the
two sublattices (e.g., $A$) formed by the carbon atoms positioned, vertically, at the centers 
(respectively, corners) of the hexagons in the adjacent layers. Within each layer, 
such a depletion (and excess occupation of the complementary sublattice, e.g., $B$)  
conforms to one of the two degenerate CDW patterns which then 
alternate between the layers, thereby keeping the electrons in the adjacent layers 
as far apart as possible and further strengthening the propensity towards the excitonic instability. 
Although the minimal strength of the inter-layer Coulomb repulsion required 
for CSB to occur in HOPG remains unknown, this whole situation, too, changes 
when the system is exposed to magnetic field normal to the layers
which promotes CSB even in the absence of the inter-layer coupling.

In the presence of the magnetic field $B$, the Dirac fermion Green function  
remains diagonal in the space of the
physical spin (not to be confused with the $4\otimes 4$ space of the $\gamma$-matrices
representing the orbital dynamics), and, for a given $\sigma$, it reads as 
\be
{\hat G}(x,y)=e^{{i\over 2}(x-y)_\mu A_\mu(x+y)}{\hat {\cal G}}(x-y)
\ee
where the translationally non-invariant phase factor contains   
the vector potential of the external field  $A_\mu(x)=(0,-Bx_2/2,Bx_1/2)$. 
Upon separating this factor out, 
one obtains a translationally (albeit, not Lorentz-) invariant Green function 
${\hat {\cal G}}(x)$.

The effect of the fermion interactions can be fully accounted for
by introducing the gap function $\Delta(p)$ as well as the wave function ($Z$) and velocity
($Z_v$)
renormalization factors into the Fourier transform of ${\hat {\cal G}}(x)$ 
given by the integral representation (hereafter, we use the units $v=e=\hbar=1$ 
and the relativistic notations, such as $p_\mu=(\epsilon, {\bf p})$)
$$
{\hat {\cal G}}(p)={i\over B}\int^\infty_0 ds 
\exp(-{s\over B}( \Delta^2(p)- Z^2\epsilon^2 + Z^2_v {\bf p}^2 {{\tanh s}\over s} ) )
$$
\be
( (\Delta(p)+Z\epsilon{\hat \gamma_0})(1-i{\hat \gamma}_1{\hat \gamma}_2\tanh s)-
 Z_v{\hat {\bf \gamma}} {\bf p} (1-{\tanh}^2s))
\ee
which the exact Green function naturally inherits from the bare one, 
${\hat {\cal G}}_{0}(p)$, given by Eq.(4) with $Z=Z_v=1$ and $\Delta(p)=0$.

Owing to its non-perturbative nature, the phenomenon of CSB   
eludes weak-coupling analysis based on perturbation theory.
Nonetheless, akin its relativistic counterpart,
the occurrence of CSB in the system of the Coulomb interacting
Dirac fermions can be inferred from the non-perturbative solution  
of the Dyson equation for the renormalized Green function   
\be
{\hat {\cal G}}^{-1}(p)-{\hat {\cal G}}^{-1}_{0}(p) = ig\int\!{d^3{k}\over (2\pi)^3}
Z{{\hat \gamma}_0{\hat {\cal G}}({p+k}){\hat \gamma}_0\over |{\bf k}|+ Ng\chi({k})} 
\ee
In Eq.(5) we made use of the Ward identity relating
the vertex function ${\hat \gamma}_0Z$ to the energy derivative of  
${\cal G}^{-1}(p)$ and cast the effective intra-layer Coulomb interaction 
in the form governed by the scalar (density-density) 
component of the fermion polarization operator 
\be
\chi(k)=i{\rm Tr}
\int {d^3{p}\over (2\pi)^3}
Z{\gamma}_0{\hat G}({p+k}){\gamma}_0{\hat G}({p})
\ee
As a result of the broken Lorentz invariance due to both, the non-relativistic 
nature of the Coulomb interaction and the presence of the magnetic field,  
the solution of the gap equation (5) can feature totally different  
dependencies on the energy and momentum variables. 
Apart from the Zeeman shift $\sigma \mu_BB$ of the  
position of the Fermi level for the spin-$\sigma$ fermions, Eq.(5) remains spin degenerate.

The analysis of Eq.(5) is complicated by the fact that, unlike in  
the previous studies of the excitonic transition in semimetals with 
overlapping conduction and valence bands, the naive picture of static Debye screening
($\chi_k\approx const$) fails to properly describe the feedback of the planar Dirac fermions 
on the bare Coulomb interaction. Instead, in the undoped or lightly doped graphite with a 
low density of carriers the zero temperature fermion
polarization (6) can only be expressed in the form of a cumbersome double integral
\cite{Farakos}
$$
\chi(k)={ Z{\bf q}^2\over 2{\sqrt{\pi B}}}
\int^\infty_0{{\sqrt{u}}du\over \sinh u}
\int^1_{-1}dv (\cosh uv-v\coth u\sinh uv)
$$
\bea
\exp(- {u\over B}( \Delta^2-
{1\over 4} (1-v^2) Z^2 \omega^2) -
\nonumber\\
-{Z^2_v{\bf q}^2\over 2B\sinh u} (\cosh u-\cosh uv) )  
\eea
Nonetheless, a progress towards obtaining the solution of Eq.(5)
can still be made in the strong field limit
where the distance betweent the adjacent Landau levels 
of the non-interacting Dirac fermions $E_n=\pm {\sqrt {2|n|B}}$
by far exceeds the Coulomb interaction-related energy gap $\Delta$,
and the only relevant fermion states appear to be those of the so-called
lowest Landau level (LLL)
with $n=0$. 
In this regard, the problem at hand bears a certain resemblance to the 
Fractional Quantum Hall effect (FQHE) in the system of non-relativistic
fermions with a parabolic dispersion.
In contrast to the spatially homogeneous 
FQHE, however, the sought solution of the gap equation corresponds to
a non-uniform CDW ground state, as explained above. 

In the strong field approximation $\Delta\ll {\sqrt B}$ 
the fermion Green function reduces to its LLL projection  
\be
{\hat {\cal G}}_{LLL}(p)\approx
ie^{-Z_v^2{\bf p}^2/B}{Z\epsilon{\hat \gamma}_0-\Delta\over \Delta^2-Z^2\epsilon^2}
(1-i{\hat \gamma}_1{\hat \gamma}_2)
\label{Gapprox}
\ee 
and, correspondingly, 
the zero temperature fermion polarization receives its main contribution from
the transitions between the LLL and the first excited Landau level  
\bea 
\chi_{LLL}(\omega,{\bf k})
\approx\sqrt{2B} {Z{\bf k}^2\over B-Z^2\omega^2/2}e^{-Z_v^2{\bf k}^2/2B} 
\label{chiapprox}
\eea
Neglecting the wave function, velocity, and vertex renormalizations 
in the scalar part of Eq.(5) 
(see below) one readily obtains a closed equation for the gap function
\bea
\Delta(p)=i\int {d\omega d{\bf k}\over (2\pi)^3}
{\Delta({k+p})\over (\epsilon+\omega+i\delta)^2-\Delta^2({k+p})}
\nonumber\\
\frac{ g e^{-({\bf (k+p)}^2+{\bf p}^2)/B} }
{ |{\bf k}|+ \sqrt{B} gN{\bf k}^2e^{-{\bf k}^2/2B}(B-\omega^2/2)^{-1} }
\eea
Eq.(10) should be contrasted with that derived in the strong field
limit in the case of $QED_3$ where the LLL projection eliminates all but the scalar
component of the Lorentz-invariant abelian gauge interaction, 
thus resulting in the gap equation (10) with the term $|{\bf k}|$ 
in the denominator of the integrand  
replaced by $k^2={\bf k}^2-\omega^2$, as required by the Lorentz invariance of the 
bare gauge field spectrum.

It is this difference which gives rise to the non-trivial energy dependence of the 
solution of the $QED_3$ gap equation found in Ref.\cite{Farakos}. By contrast, 
in our case the solution of Eq.(10) remains independent of the energy
variable as long as $\epsilon\lesssim \sqrt{B}$ and, being a sole function of the momentum, 
it falls off faster than $e^{-2{\bf p}^2/B}$ for $|{\bf p}| > \sqrt{B}$. 
It can be also shown that the approximate constancy of $\Delta({p})$ at small momenta
and energies justifies our neglecting the wave function and velocity
renormalization ($Z=Z_v=1$).

The magnitude of the zero-momentum gap $\Delta=\Delta(0)$ estimated from Eq.(10)   
\be
\Delta\approx {\sqrt{B}\ln (Ng)\over 4\pi N}
\ee
demonstrates that the strong field condition holds at large $N$ 
regardless of the field strength $B$ (which should, however, be smaller
than the width of the linear part of the bare quasiparticle spectrum).
In the physical case $N=2$, a numerical analysis which   
does not rely on the LLL approximation \cite{Leal} shows a qualitatively similar  
behavior, thus supporting the conclusions drawn in the large-$N$ limit.

At finite temperatures the gap decreases,
and its momentum dependence at $|{\bf p}|\lesssim (T\sqrt{B})^{1/2}$ 
becomes even less pronounced. Although the prohibitive form of the finite temperature 
fermion polarization impedes analytical calculations, 
its main effect can be taken into account 
through the modified lower cutoff in the logarithmically   
divergent momentum integral in Eq.(10) which now becomes 
the larger of $\sqrt{B}/gN$ and $(T\sqrt{B})^{1/2}$.

After being extended to finite temperatures, Eq.(10) yields
the self-consistent equation for the magnitude of the gap at small momenta 
\be
\Delta_{T}\approx {\sqrt{B}\over 4\pi 
N}\ln{\sqrt{B}\over max(\sqrt{B}/gN; (T\sqrt{B})^{1/2})}\tanh{\Delta_{T}\over 2T}
\ee
Eq.(12) possesses a non-trivial solution 
below the transition temperature $T_c(B)$ whose large-$N$ estimate is
given by the expression 
\be
T_c\approx {\sqrt{B}\ln N\over 16\pi N}
\ee
Conversely, Eq.(13) determines a threshold magnetic field $B_c(T)\propto T^2$
which has to be exceeded in order for CSB to occur at a non-zero temperature $T$.
This relation defines  
a critical line in the $B-T$ phase diagram, along which the gap vanishes as 
\be
\Delta_{T}(B\to B_c)\propto \sqrt {B-B_c(T)}
\ee
As far as the nature of the CSB transition is concerned, the dependence (14)
is characteristic of the second order transition, whereas the conjectured zero field 
transition (for $N<N_c$) is of the topological (Kosterlitz-Thouless) kind \cite{Leal}. 

These predictions should be compared with the available experimental evidence
obtained from the HOPG samples showing metallic behavior of the zero field resistivity. 
The data of Ref.\cite{Kempa} indicate that, albeit absent in zero field,
the apparent semimetal-insulator transition can be 
induced by magnetic field normal to the layers.
The insulating behavior was found to set in at the field-dependent characteristic 
temperature fitted as $T^*\propto\sqrt{B-B_0}$ which, apart from the offset field
$B_0$, agrees with Eq.(13) and   
falls into the experimental range $\lesssim 100K$ for the applied magnetic field
$B\lesssim 0.2 T$ (in order to avoid confusion we remind that in our quasi-relativistic 
system the role of the "speed of light" is played by $v$). 
The proposed orbital (as opposed to the spin-related) nature of the observed phenomenon is
also consistent with the findings of Ref.\cite{Kempa}, according to which 
the in-plane magnetic field has a substantially (two orders of magnitude) weaker effect. 

In fact, the spin degeneracy between the triplet and singlet excitonic gaps (for $N=2$) 
will be lifted upon including the short-ranged Coulomb exchange interaction omitted in Eq.(2)
which involves transitions between the conduction and valence bands. Alongside
with the Zeeman coupling, the latter favors the triplet excitonic order parameter, 
in accord with the Hund's rule.
In the doped system, the 
spin up- and down- states will then be occupied asymmetrically at low temperatures, resulting in  
the occurrence of a ferromagnetic spin polarization ${\bf M}=Tr({\vec {\bf \sigma}}G)$ 
in a window of the electron chemical potential $\mu$ set by the gap $\Delta$. 
The induced ferromagnetic
moment will then be proportional to $\mu\sim\Delta$ as well,
which compares favorably with the electron spin resonance (ESR) data in HOPG  
showing the presence of the magnetization ${\bf M}\propto\sqrt{B-B_0}$ on the 
insulator side of the field-induced transition
\cite{Sercheli}.

Before concluding, we comment on the previous attempts to apply the alternate  
scenario of the magnetic field-driven CSB in the Lorentz-invariant HY model 
to the analysis of the quasiparticle 
transport in the mixed state of the planar $d$-wave superconductors \cite{Liu}.
Specifically, the authors of Ref.\cite{Liu} focussed on the experimentally
observed kink-like feature in the magnetic field dependence of the
total (inclusive of both the normal quasiparticle and phonon
contributions) thermal conductivity, 
the position of the kink scaling with temperature as $B^*(T)\propto T^2$ \cite{Ong}
(this behavior was not seen, however, in a more recent experiment \cite{Ando}).

On the theoretical side, 
the suggestion of using the standard HY model to describe 
the nodal $d$-wave quasiparticles in the mixed state must be taken cautiously.
Apart from the default choice of the quasiparticle interaction
in the form of the attractive HY coupling, 
the "magnetic catalysis" scenario of Refs.\cite{Liu}
requires that the total effective magnetic field, which the nodal quasiparticles are exposed 
to, has a non-zero overall flux, resulting in the formation of the Landau levels.

However, as pointed out by several authors
\cite{Franz}, the correct picture of the $d$-wave quasiparticle spectrum 
is rather that of the extended energy bands, owing to the fact that in the mixed state the
quasiparticles are actually experiencing both, the external magnetic field
and the solenoidal superfluid flow created by the vortices. On average, 
the two fluxes exactly cancel each other \cite{Franz}, thus rendering inapplicable the  
standard HY mechanism of the "magnetic catalysis" even in the regime of
weak-to-moderate magnetic fields where the individual vortices
strongly overlap and the physical magnetic field is almost uniform.
In the light of the above, HOPG might be the only presently known example 
of a condensed matter system where the phenomenon of "magnetic catalysis" can 
indeed occur.

In summary, we study the problem of the Coulomb interaction-driven electronic instabilities 
in layered graphite in the presence of magnetic field.   
Elaborating on the relativistic-like description of the low-energy
quasiparticle excitations in a single sheet of graphite, 
we propose a possible explanation for the recently discovered 
field-induced semimetal-insulator transition
by showing that applied magnetic field 
induces the excitonic insulator phase, thus gapping up the quasiparticle spectrum
and creating a site-centered CDW.

Since the phenomenon in question was only observed in  
perpendicular magnetic field, we believe that the experimental findings of 
Refs.\cite{Kempa,Sercheli}  reveal some novel, intrinsically two-dimensional, physics 
whose origin is different from that of the (semi)metal-insulator transitions  
observed in other carbon-based materials, such as the one-dimensional
($(Ru,Cs)C_{60}$) as well as three-dimensional
($KC_{60}$) alkali doped fullerides. The latter are likely to be associated 
with the structural instabilities and/or accompanied by 
the onset of the antiferromagnetic spin-Peierls state. 
In order to decisively discriminate between these possibilities, such experimental techniques
as ESR, NMR, x-ray diffraction, and electron photoemission, can be further 
implemented.
    
The author is grateful to Y. Kopelevich for communicating his experimental results prior
to publication and S. Washburn for a valuable discussion. 
This research was supported by the NSF under Grant No. DMR-0071362.

\wt
\end{document}